# Information hiding cameras: optical concealment of object information into ordinary images

Bijie Bai [†,1,2], Ryan Lee[†,3], Yuhang Li[1,2], Tianyi Gan[1,2], Yuntian Wang[1,2], Mona Jarrahi[1,2], and Aydogan Ozcan[*,1,2,4]

[1]Electrical and Computer Engineering Department, University of California, Los Angeles, CA, 90095, USA.

[2]California NanoSystems Institute (CNSI), University of California, Los Angeles, CA, USA.

[3]Computer Science Department, University of California, Los Angeles, 90095, USA.

[4]Bioengineering Department, University of California, Los Angeles, 90095, USA.

[†]Equal contributing authors

[*]Correspondence: Aydogan Ozcan. Email: ozcan@ucla.edu

## Abstract

The security of sensitive information is crucial. Data protection methods like cryptography, despite being effective, inadvertently signal the presence of secret communication, thereby drawing undue attention. These digital methods also require substantial computational resources and struggle with computing time/speed when handling large data. Here, we introduce an optical information hiding camera integrated with an electronic decoder, optimized jointly through deep learning. This information hiding–decoding system employs a diffractive optical processor as its front-end, which transforms and hides input images in the form of ordinary-looking patterns that deceive/mislead human observers. This information hiding transformation is valid for infinitely many combinations of secret messages, all of which are transformed into ordinary-looking output patterns, achieved all-optically through passive light-matter interactions within the optical processor. By processing these ordinary-looking output images, a jointly-trained electronic decoder neural network accurately reconstructs the original information hidden within the deceptive output pattern. We numerically demonstrated our approach by designing an information hiding diffractive camera along with a jointly-optimized convolutional decoder neural network. The efficacy of this system was demonstrated under various lighting conditions and noise levels, showing its robustness. We further extended this information hiding camera to multi-spectral operation, allowing the concealment and decoding of multiple images at different wavelengths, all performed simultaneously in a single feed-forward operation. The feasibility of our framework was also demonstrated experimentally using THz radiation. This optical encoder–electronic decoder-based co-design provides a novel information hiding camera interface that is both high-speed and energy-efficient, offering an intriguing solution for visual information security.

## Introduction

In an era where digital communication permeates every aspect of life, the security of sensitive information has become increasingly vital[1–4]. One approach for data protection is based on cryptography[5–7], which secures messages through encryption, making them accessible only to intended recipients. While cryptography is a well-established field, it makes use of encrypted messages, which can draw attention when perceived, thereby signaling the existence of secret communication between the sender and the recipient. This raises the need for solutions that can hide sensitive data behind ordinary-looking patterns or images without arousing suspicion. Building upon this need, methods of steganography have emerged to conceal sensitive information within ordinary-looking objects or media, masking the presence of confidential data and avoiding unwanted attention[8,9]. For instance, the traditional Least Significant Bit (LSB) substitution[10–13] method digitally modifies the least noticeable bits of the pixel values in an image to embed secret data, making changes that are invisible to the human eye. Another digital approach involves the use of transform-domain techniques, where the confidential data are embedded in frequency components of the image, utilizing digital algorithms based on e.g., discrete cosine transforms or wavelets[9,14–16]. However, these traditional methods of digital steganography, though effective in hiding data, can potentially suffer from limitations such as limited data embedding capacity, vulnerability to image compression and noise, and trade-offs between data security and computational complexity[9,17,18]. Alternative techniques, such as adaptive steganography[19–21] and deep learning-based steganography[22–26] offer more secure solutions, which, however, can be computationally intensive and demand active digital computing resources for both information hiding and recovery. In general, there is a growing need for alternative approaches that combine high speed, energy efficiency, as well as robustness and versatility for private information concealment.

In this work, we report an all-optical information hiding camera design (Fig. 1) that uses programmed light diffraction through a passive structured material to optically transform input messages into ordinary-looking (and misleading) output patterns that can be decoded using a neural network back-end. The all-optical information hiding diffractive processor and the electronic decoder are jointly trained using a deep learning-based optimization approach, and this collaborative framework maintains its functionality for all combinations of possible input messages presented at the input field of view (FOV). We numerically demonstrated this approach using MNIST handwritten digits[27] and jointly optimized a five-layer information hiding diffractive processor and a decoding convolutional neural network (CNN). The blind testing results proved the feasibility of this framework, with the diffractive processor successfully transforming arbitrarily selected handwritten input digits, never seen before, into a uniform, ordinary-looking digit "8" and the electronic neural network successfully reconstructing the original handwritten digits that were placed at the input. The system's resilience was tested under various noise levels and different lighting conditions, including spatially uniform and non-uniform illuminations, confirming its robustness in performance. Furthermore, we numerically demonstrated a multi-spectral information hiding and decoding scheme, where the information hiding camera concealed multiple independent information channels at its input FOV, each carried by a distinct illumination wavelength, into ordinary-looking colorful output images, from which the decoding CNN subsequently reconstructed the original multi-channel messages. Finally, we experimentally validated the feasibility of this approach by designing and fabricating an information hiding camera operating at the terahertz (THz) part of the spectrum. The measurements of the camera's output, revealing misleading ordinary-looking images, were successfully processed by the jointly trained CNN decoder to recover the concealed input messages with high fidelity.

Integrating a deep learning-based optical-electronic co-design strategy, our approach represents a novel solution for visual data privacy and secure sharing of information. Compared to various digital information hiding methods, optical information concealment through a physical diffractive processor offers a faster, more energy-efficient and scalable solution.

## Results

### Optical information concealment

We first numerically demonstrate our presented framework of optical information concealment using the MNIST handwritten digit dataset[27]. As shown in Fig. 1, our information hiding – decoding system consists of two connected components: (1) an all-optical diffractive processor that conceals any arbitrary input images and transforms them into misleading, ordinary-looking patterns; and (2) a subsequent electronic decoding network that communicates with the diffractive processor to accurately reconstruct the original information hidden within the misleading patterns.

The diffractive processor, as illustrated in Fig. 2a, is designed to conceal the input information all-optically and operates at a spatially coherent illumination wavelength of $\lambda$. It consists of five diffractive layers that axially span $<110\lambda$ in total, where each layer contains 120×120 phase-modulating diffractive features with a lateral size of $\sim\lambda/2$ per phase feature. The electronic decoding network is a convolutional neural network (CNN) (see the Methods section). In this setup, an arbitrarily selected input MNIST image is first optically processed by the diffractive camera, resulting in an information-hiding output image that misleads human observers as if a different object was imaged by the camera. Then, the resulting intensity image at the output of the diffractive camera is normalized to 1 and fed into the jointly optimized electronic decoding CNN for the reconstruction of the actual input MNIST image.

For our demonstration, we selected a random MNIST digit "8" as our information hiding "dummy" message to be consistently generated by the diffractive processor for all possible handwritten input digits. For example, a sequence of handwritten messages "*0, 1, 2, 3, 4, 5, 6, 7, 8, 9*" will be all-optically converted to the same handwritten digit "8", and will yield, at the output of the diffractive camera, a sequence of "*8, 8, 8, 8, 8, 8, 8, 8, 8, 8*". This misleading output sequence of "*8, 8, 8, 8, 8, 8, 8, 8, 8, 8*" will then be processed by the electronic back-end decoder to recover the actual input messages "*0, 1, 2, 3, 4, 5, 6, 7, 8, 9*". This information hiding camera and the electronic decoding CNN were trained jointly using the stochastic gradient descent algorithm. During the iterative training process, the phase modulation values of the diffractive camera layers and the weights of the convolutional layers in the CNN were simultaneously updated based on a customized loss function (detailed in the Methods section). This loss function is designed to achieve two aims: (1) enhance the structural similarity between the diffractive camera's output image and the selected information-hiding "dummy" digit "8", thereby helping the diffractive camera to convert any input handwritten digit images into ordinary-looking output images, each closely resembling the selected handwritten digit "8"; and (2) improve the resemblance of the CNN output to the original input image of the system, ensuring accurate reconstruction of the original input information from the misleading "dummy" patterns.

After this joint training, we conducted a numerical evaluation using 10,000 handwritten digits from the MNIST test dataset that were never used during the training phase. As shown in Fig. 2b, the diffractive processor successfully transformed various input handwritten digits into uniform, ordinary-looking handwritten "8" patterns (images shown after normalization), effectively concealing the original information and misleading human observers. The electronic network successfully decoded the concealed information, accurately reconstructing the original input images of the handwritten digits from the optically transformed patterns of handwritten "8".

For the results presented in Fig. 2, a normalization step for the diffractive camera's output images was incorporated before the hidden information was decoded by the CNN, which can be considered analogous to the auto-exposure function in practical camera systems. However, this normalization step is not necessary for the operation of our information hiding camera design. To showcase this, we developed an additional diffractive model where this image normalization step was omitted. The results of this diffractive camera are summarized in Supplementary Fig. 1, which demonstrates a similar level of success. The information hiding camera accurately converted various input handwritten digits into misleading, ordinary-looking patterns of handwritten "8" images, where the jointly trained CNN decoder faithfully retrieved the hidden handwritten digit information that was originally at the input. Additional examples of blind testing results for both the normalized and non-normalized diffractive camera models are presented in Supplementary Videos 1 and 2. These successful numerical demonstrations validate the efficacy of our all-optical information hiding camera with an electronic decoder used for information retrieval.

**Evaluation of information hiding diffractive camera performance under varying illumination conditions**

In addition to demonstrating the primary functionality of our information hiding camera framework, we also evaluated the system's resilience under different illumination conditions. To assess this, we first conducted blind inferences by testing the trained diffractive camera model (illustrated in Fig. 2) under varying illumination intensities. As shown in Fig. 3a, despite changes in the input image brightness, the information hiding diffractive camera consistently produced output patterns resembling the ordinary handwritten digit "8", successfully concealing the original information; subsequently, the CNN decoder reconstructed the original input images with high fidelity under all these varying conditions. To quantitatively assess the system's performance, we also computed the Pearson Correlation Coefficient (PCC)[28] between the original input images and the CNN reconstructions across varying illumination levels, evaluating their pixel-wise linear correlation. The PCC value between two given images A and B is defined as:

$$\text{PCC}(A, B) = \frac{\sum (A - \bar{A})(B - \bar{B})}{\sqrt{\sum (A - \bar{A})^2 \sum (B - \bar{B})^2}} \tag{1}$$

where $\bar{A}$ and $\bar{B}$ are the average intensity values of the images $A$ and $B$, respectively. The line plot in Fig. 3a confirms that the CNN reconstruction PCC value remains stably high at 0.99 across different testing illumination levels, effectively preserving the structure and contrast details of the original image at the input of the diffractive camera. Additional blind testing examples under different illumination intensities are provided in Supplementary Video 3.

We also tested the same information hiding – decoding system with input objects under *spatially non-uniform* illumination patterns. In this test, the illumination level of one quadrant of the input objects was purposely made different (dimmer or brighter) from the rest, creating scenarios where the input objects are under non-uniform lighting conditions. Despite these illumination pattern variations that were not represented during the training of the system, our information concealing – decoding framework, as illustrated in Fig. 3b, effectively maintained its functionality. The information hiding camera output consistently displayed the misleading patterns of an ordinary handwritten digit "8", from which the CNN decoder accurately reconstructed the original input information, achieving a PCC value greater than 0.94 within our test images. Additional blind testing examples are also provided in Supplementary Video 4.

The same set of illumination analyses was also conducted for the information hiding diffractive camera model trained *without* the normalization of the camera's output images (illustrated in Supplementary Fig. 1). In the case of uniformly varying illuminations, as demonstrated in Fig. 4a, the brightness of the diffractive camera output scaled proportionally with the input illumination level, yielding dimmer to brighter output images as a function of the illumination intensity. Correspondingly, the reconstruction brightness from the electronic decoder also varied. Although the reconstruction PCC values of the electronic decoder degraded as the illumination deviated from the training level, they remained consistently above 0.9, indicating a robust performance under these varying test illumination conditions (see Fig. 4a). Furthermore, the same spatially non-uniform illumination conditions were also evaluated for the model trained without the output image normalization. As depicted in Fig. 4b, when subjected to non-uniform lighting conditions, the diffractive processor output exhibited slight variations in brightness levels. Despite this, the electronic decoder managed to reconstruct the original input images with an average PCC value above 0.94.

These studies collectively demonstrate the adaptability and robustness of our all-optical information hiding and electronic decoding framework in handling varying complex lighting scenarios. Whether facing uniformly varying or spatially non-uniform illumination conditions, the system can maintain its core functionality of information concealment and high-fidelity image reconstruction, even though the original model was not trained with such varying illumination conditions. This resilience highlights the adaptability of our system under different applications where the lighting conditions can be unpredictable and diverse.

**Evaluation of model performance under noise**

Another essential aspect of assessing the robustness of our information hiding system involves evaluating its performance in the presence of noise. Specifically, we focused on the system's behavior when the all-optical diffractive processor is subjected to Gaussian noise[29,30]. To conduct this evaluation, we first took the already trained model reported in Fig. 2 and added random Gaussian noise ($\varepsilon$) onto the diffractive processor's raw output images. The added Gaussian noise followed a normal distribution, i.e.,

$$\varepsilon \sim \mathcal{N}(\mu = 0, \sigma_{\text{te}}^2) \qquad (2)$$

Following the addition of random noise, any negative values in the resulting image were clipped to zero, and the image was then normalized before being fed into the electronic decoder CNN for the reconstruction of the hidden message. As shown in Fig. 5a, the system was tested under various levels of noise, and a degradation in the CNN reconstruction fidelity was observed as the testing noise level increased from $\sigma_{te}^2 = 5 \times 10^{-4}$ to $\sigma_{te}^2 = 5 \times 10^{-3}$.

In addition to testing with post-training noise introduction, we also evaluated the system's performance when trained with noise. For this evaluation, a separate model was trained with Gaussian noise added to the diffractive processor's output, aiming to "vaccinate" the model[31] against noise by exposing it to various noisy conditions during the training process. Figure 5b shows the results of the model trained with a vaccination noise level of $\sigma_{tr}^2 = 5 \times 10^{-2}$ and tested under varying noise levels ($\sigma_{te}^2$). It is demonstrated that the electronic decoder maintained high reconstruction quality at a testing noise level as high as $\sigma_{te}^2 = 5 \times 10^{-2}$, with a gradual decrease in its performance as the noise level further increased to $\sigma_{te}^2 = 1 \times 10^{-1}$. Supplementary Video 5 further provides a side-by-side comparison of the models trained *with* and *without* noise vaccination, tested across the same spectrum of varying noise levels. This comparison highlights the enhanced resilience of the vaccinated model against more substantial noise perturbations.

Further expanding our exploration, we also trained two additional hybrid models with vaccination noise levels set at $\sigma_{tr}^2 = 5 \times 10^{-3}$ and $\sigma_{tr}^2 = 1.5 \times 10^{-1}$, respectively. Then, we analyzed the CNN reconstruction PCC values and the diffractive output energy efficiency metrics for the three vaccinated diffractive models along with the original non-vaccinated model. These models were evaluated across testing noise levels ranging from $\sigma_{te}^2 = 0$ to $\sigma_{te}^2 = 3 \times 10^{-1}$ using 10,000 MNIST testing images, and the average quantitative results are summarized as line plots in Fig. 5c. When observing the CNN reconstruction fidelity (Fig. 5c, left panel), the non-vaccinated model (red curve) is more vulnerable to noise at the diffractive camera, exhibiting a drastic drop in its PCC values at relatively lower noise levels. In contrast, the vaccinated models demonstrate enhanced resilience to noise and maintain a relatively high PCC across a broader range of noise levels, albeit with a slight reduction in noise-free ($\sigma_{te}^2 = 0$) image reconstruction fidelity. Furthermore, we observed that the models trained with higher levels of noise vaccination exhibit greater resistance to noise perturbations. They retain high PCC values for testing noise levels that are below their training vaccination level (i.e., when $\sigma_{te}^2 < \sigma_{tr}^2$), whereas a more significant performance degradation is observed when the testing noise level surpasses the training vaccination noise level ($\sigma_{te}^2 > \sigma_{tr}^2$).

The output energy efficiency of the information hiding diffractive camera (shown in Fig. 5c, right panel) is calculated as the ratio of the total diffractive processor output energy to its total input energy prior to the introduction of noise and normalization, i.e.,

$$\boldsymbol{\eta}(I, O_D) = \frac{\sum O_D}{\sum I} \qquad (3)$$

where $I$ denotes the input image intensity and $O_D$ denotes the raw output intensity image of the information hiding diffractive camera. As shown in Fig. 5c, a notable *increase* in the output diffraction efficiency is observed with higher levels of noise vaccination in training. For instance, the output diffraction efficiency of the non-vaccinated information hiding camera model is 8.8%, which rises to **31.9%**, **53.7%**, and **89.1%** for the models trained with progressively higher levels of noise vaccination (see Fig. 5c, right panel). This increase suggests that the diffractive processor strategically optimized its output energy efficiency to preserve a high signal-to-noise ratio (SNR) to confront the elevated noise levels at the output. These analyses collectively demonstrate that vaccinated training can help us achieve robust and reliable performance under noisy conditions that are inevitable in practical settings.

**Multi-spectral optical information concealment**

To further demonstrate the versatility of our presented framework, we next developed a multi-spectral information concealing – decoding system on the basis of a broadband diffractive optical processor. As illustrated in Fig. 6a, this multi-spectral system operates at three distinct wavelengths: $\lambda_1 = 630$ nm (red), $\lambda_2 = 530$ nm (green), and $\lambda_3 = 450$ nm (blue), aligning with the primary colors of conventional vision and display systems. Correspondingly, the input of the diffractive processor is a three-channel image, with each channel embedding a unique MNIST image associated with its respective wavelength, allowing simultaneous concealment of multiple messages/images through different wavelengths in a single optical inference step. A five-layer diffractive processor is employed to transform any arbitrarily selected multi-spectral input image into a three-channel, colorful version of an MNIST object (i.e., the handwritten digit "8" in our demonstration), which can be captured in a single snapshot by a conventional RGB sensor placed at the diffractive camera's output FOV. Following the RGB image capture at output FOV, an electronic decoder network reconstructs the original multi-channel input image using the misleading "dummy" RGB output image. Each diffractive layer of the diffractive processor has 120×120 trainable phase-modulating diffractive features, each with a lateral size of ~0.44$\lambda_3$ and the diffractive layers axially span <180$\lambda_3$ in total. The electronic CNN decoder utilizes a similar architecture as depicted in Fig. 2a, with modifications in its first and last convolutional layers to adapt to the three-channel (RGB) input/output images. The all-optical information hiding diffractive camera and the electronic decoding

CNN were jointly trained to ensure that the diffractive processor output closely mimics the geometric structure of the selected "dummy" image (a handwritten digit "8"), and that the CNN accurately reconstructs the original input information (see the Methods section for details). After this joint training, the resulting hybrid model was numerically tested using 10,000 examples, where each channel of the RGB input image was independently sampled from the MNIST testing dataset. As shown in Fig. 6b, the diffractive processor successfully concealed the original RGB image information, generating colorful and ordinary-looking output images of a handwritten "8". The subsequent processing by the CNN decoder revealed a high-fidelity reconstruction of the original multi-channel images, demonstrating the system's capability to process complex colorful image data.

Next, we conducted an analysis of the diffractive processor's energy efficiency across each illumination wavelength for all the 10,000 blind testing examples. As shown in Fig. 7a histogram, the diffractive processor tends to have higher diffraction efficiency at the green wavelength (i.e., $\lambda_2 = 530$ nm), with an average diffraction efficiency of 1.67% compared to 0.98% for red and 0.36% for blue wavelengths. This difference among the illumination wavelengths is also visually evident in the examples presented in Fig. 7b, where 64 randomly sampled output images of the multi-spectral diffractive processor predominantly displayed greenish colors at the output FOV. In addition, we also examined the correlation between the colors of the input and output images of the information hiding diffractive processor. To do this, we first converted the RGB images into HSV domain, and plotted the average hue of each input-output image pair across all the 10,000 test images. The scatter plot shown in Fig. 7c reveals an approximately linear relationship between the input and output hues, suggesting that the intensity of each output color channel is closely linked to its corresponding input color channel.

In addition to this standard multi-spectral diffractive camera model described above, we also developed an improved model, incorporating an additional training loss function to *equalize* the energy efficiency across all three wavelengths (see the Methods section). After the deep learning-based training, the resulting diffractive camera model was numerically tested using 10,000 examples, with each channel independently sampled from the MNIST testing dataset. The blind testing examples shown in Fig. 6c once again demonstrate the success of this approach, where the information hiding diffractive camera concealed the input images and generated colorful ordinary-looking output patterns, while the CNN decoder faithfully reconstructed the original RGB image information. Analyses of the output diffraction efficiency of this new model revealed that the previous dominance of the green wavelength at the output was eliminated by this additional loss term. This is supported by the fact that the histograms of the three wavelength channels closely align (Fig. 7d), and the random testing examples present a more balanced color outcome at the diffractive processor output FOV (Fig. 7e). Moreover, the scatter plot in Fig. 7f, depicting the input-output hue relationship, not only maintains an approximately linear relationship, but also spans a broader output hue range compared to the standard model trained without the energy efficiency equalization loss term. These observations further prove the success of the equalization loss term, broadening the output color space achievable by the diffractive optical processor.

**Experimental demonstration of the information hiding diffractive camera system**

We demonstrated an experimental implementation of our information hiding camera system using terahertz radiation. For this experiment, we first trained a hybrid model using a single-layer diffractive processor for information concealment and a subsequent electronic CNN for decoding the concealed input information (Fig. 8a). Following the earlier numerical demonstrations, the diffractive processor was optimized to hide the input information (handwritten digits) and convert the input images into ordinary-looking images of a handwritten digit "8". The diffractive processor was trained for monochromatic illumination at $\lambda = 0.75$ mm, and its diffractive layer consisted of 100×100 trainable diffractive features, each with a lateral size of ~$0.67\lambda$. The electronic CNN follows the same architecture as depicted in Fig. 2a, which was jointly optimized with the all-optical information hiding camera to reconstruct the original inputs. Throughout the training phase, the same vaccination strategy as described in the noise evaluation section earlier was implemented to overcome potential detection noise issues during the experiments; for this vaccination, a level of $\sigma_{\mathrm{tr}}^2 = 5 \times 10^{-2}$ Gaussian noise was employed during our training. After the training, the resulting diffractive layer was fabricated using 3D printing (see Fig. 8b), creating a physical information hiding camera. This assembled information hiding camera was experimentally tested under a scanning THz system operating, as shown in Fig. 8c. The output FOV of the diffractive processor was scanned using a THz detector to generate the output "dummy" images.

For the blind testing of our fabricated diffractive processor, we selected 10 different MNIST testing digits that were never used in the training. The measured output images of the diffractive processor were then normalized and processed by the CNN decoder for reconstruction of the concealed information. Figure 8d summarizes the experimental measurements from our 3D-printed diffractive processor and the corresponding reconstructions from the CNN decoder, alongside the numerical simulation results for each test object. The experimental outcomes from both the information hiding diffractive processor and the decoder CNN demonstrate a high correlation with the numerically expected results. The information hiding diffractive camera successfully converted various input test images into an ordinary-looking handwritten digit "8", whereas the electronic CNN decoder faithfully restored the original input information from these misleading "dummy" output images. The success of this proof-of-concept experiment further validates the feasibility of our information hiding – decoding system.

## Discussion

In this work, we demonstrated a diffractive framework for all-optical information hiding with electronic decoding, which has the potential to enhance data privacy in digital communications. By developing an all-optical information hiding diffractive camera that transforms input images into deceptive, misleading patterns, with an electronic decoder designed to retrieve the original data, we demonstrated how all-optical diffractive processors can offer a secure and efficient alternative to existing digital information hiding methods. The system's performance was successfully tested under various illumination patterns and different wavelengths, also showing resilience against output detection noise and potential experimental misalignments.

This information hiding diffractive camera system operates by transforming desired secure messages into seemingly regular, ordinary-looking images or patterns, effectively deceiving unauthorized observers while securely transmitting information to the intended recipients who have access to the accurate decoder network. This framework also has the potential for multi-user secret communication, where a single front-end information hiding diffractive camera can be integrated with multiple distinct CNN decoders, each recovering only specific parts of the concealed messages, for example, based on data-class specific transformations[32–34]. By assigning a unique decoder to each group of users based on their authorization levels, the system can ensure that individuals can only access the messages specifically intended for them. In addition to these, the system can be further enhanced by designing the decoder CNN to output a transformed/encrypted version of the original image (as opposed to outputting the original input image as we demonstrated here), which can be used as an additional layer of security for sharing hidden information.

Although the presented results used spatially coherent illumination, information hiding diffractive camera designs can also be extended to spatially incoherent illumination schemes[35,36]. Under spatially incoherent illumination, the spatially varying intensity point spread function (PSF) of a diffractive processor, i.e., $P$, can be written as $P(x, y; x', y') = |p(x, y; x', y')|^2$ where $p(x, y; x', y')$ denotes the spatially-varying coherent PSF of the same diffractive processor between its input and output apertures defined by the coordinates $(x', y')$ and $(x, y)$, respectively[35]. It has been shown that, under spatially incoherent illumination, diffractive optical processors can accurately approximate arbitrarily selected linear intensity transformations between an input and an output aperture[35,36]. Therefore, by leveraging data-driven optimization techniques, our information hiding camera framework can also be designed to adapt to spatially and temporally incoherent or partially-coherent illumination conditions, including, e.g., natural light or light-emitting diodes.

Looking forward, the applications of diffractive visual information processors in privacy-aware technologies present promising future research and development opportunities. Their attributes of high-speed information processing, energy efficiency, and scalability make diffractive processors an attractive solution for real-time secure communication of visual information and data protection systems. With the recent advancements in nanofabrication techniques, such as two-photon polymerization[37], this framework can also be scaled down to the visible spectrum, operating under spatially and temporally incoherent radiation. The success of this hybrid approach also encourages further exploration into the integration of optical and electronic computing systems for more efficient and versatile visual data processing systems.

## Methods

### Optical forward model of the information hiding diffractive camera system

The optical forward model of a diffractive processor can, in general, be characterized by two successive operations: (1) free-space propagation of the light field between two consecutive planes, and (2) the modulation of the incoming light field by each individual diffractive layer. The free-space propagation of the light field is modeled using the angular spectrum approach[38], which is expressed as:

$$u(x, y, z + d) = \mathcal{F}^{-1}\{\mathcal{F}\{u(x, y, z)\} \cdot H(f_x, f_y; d)\} \tag{4}$$

where $u(x, y, z)$ is the original light field at the coordinate of $z$ along the optical axis, and $u(x, y, z + d)$ is the resulting light field at the coordinate of $z + d$ after the free-space propagation over an axial distance of $d$. $f_x$ and $f_y$ denote the spatial frequencies along the $x$ and $y$ axes, respectively. $\mathcal{F}$ and $\mathcal{F}^{-1}$ are the 2D Fourier transform and 2D inverse Fourier transform, respectively. $H(f_x, f_y; d)$ describes the free-space transfer function, defined as:

$$H(f_x, f_y; d) = \begin{cases} \exp\left\{jkd\sqrt{1 - \left(\frac{2\pi f_x}{k}\right)^2 - \left(\frac{2\pi f_y}{k}\right)^2}\right\}, & f_x^2 + f_y^2 < \frac{1}{\lambda^2} \\ 0, & f_x^2 + f_y^2 \geq \frac{1}{\lambda^2} \end{cases} \tag{5}$$

where $j = \sqrt{-1}$. $\lambda$ is the wavelength of the illumination light and $k = \frac{2\pi}{\lambda}$. The diffractive layers in this work are treated as thin optical elements, modulating the phase of the transmitted optical field. At the $l^{th}$ diffractive layer, the transmission coefficient of the diffractive feature located at the lateral $(x, y)$ position is described by:

$$t^l(x, y) = \exp\{j\phi^l(x, y)\} \tag{6}$$

where $\phi^l(x, y)$ is the phase modulation value of the corresponding diffractive feature.

**Experimental demonstration**

After training, the resulting diffractive layers' phase values were converted into height maps based on the refractive index of the 3D printing material. The diffractive layers, the test objects, and the holder for the mechanical assembly (see Fig. 8c) were 3D printed using Stratasys Objet30 Pro. The test objects were coated with aluminum foil to define the transmission regions. After its fabrication, the physical diffractive camera was tested using a THz continuous wave system[32,33] with $\lambda = 0.75$ mm. A digital 2×2 binning was applied after the measurement to match the object feature size used in the numerical simulations.

**Supplementary Information** includes:

- **Electronic decoding neural network architecture**
- **Training loss functions**
- **Parameters and digital implementation for training**
- **Supplementary Figure 1**
- **Supplementary Movies 1-5**

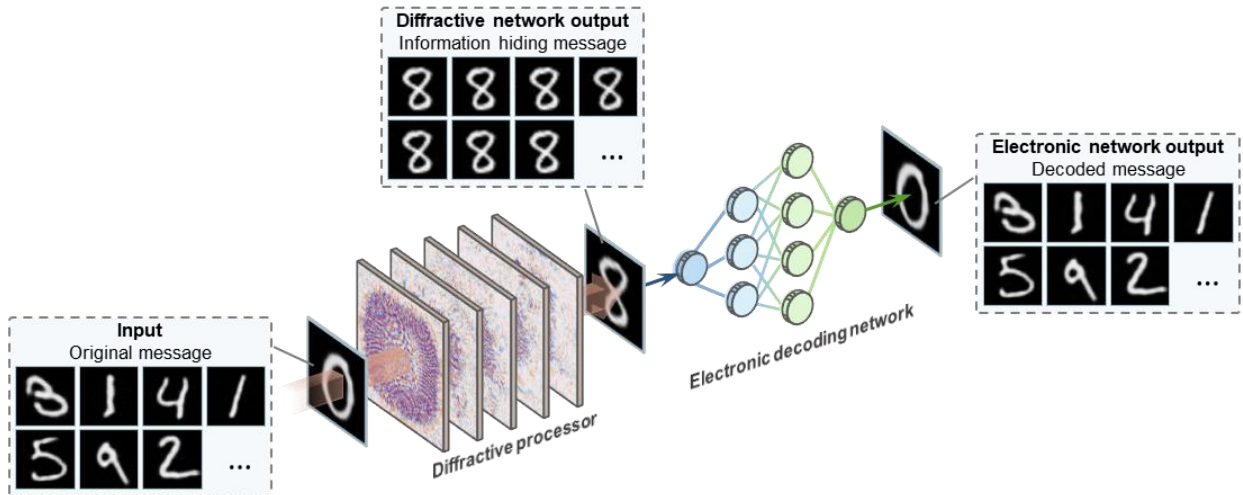

**Figure 1. All-optical information hiding camera with electronic decoding**. The diffractive processor all-optically conceals arbitrarily selected input images and transforms them into ordinary-looking "dummy" patterns that mislead human observers, whereas a following electronic decoder recovers the original image from these information hiding output messages.

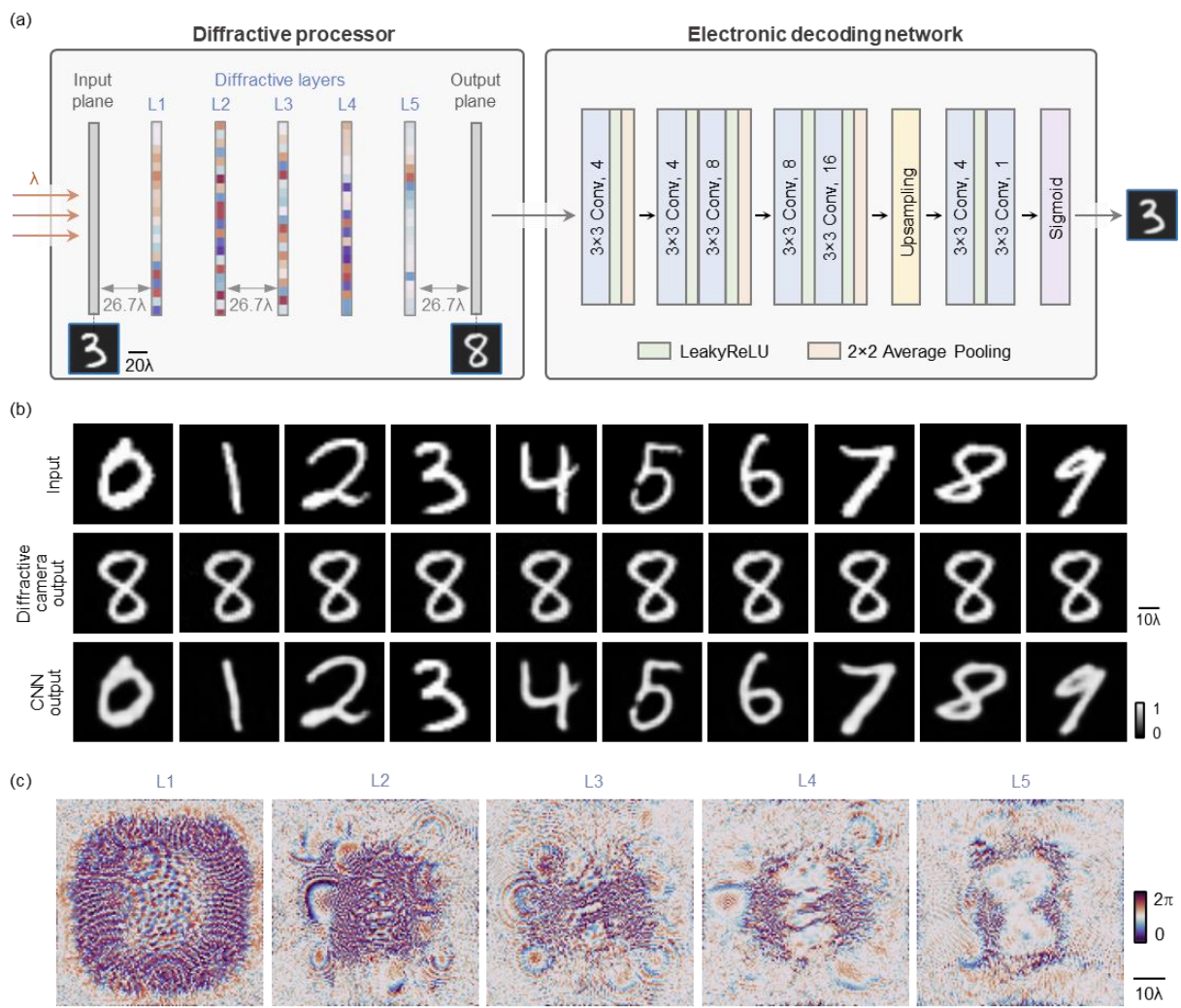

**Figure 2. Design schematic and blind testing results of the information hiding – decoding system.** (a) Design schematics of a five-layer information hiding diffractive processor and an electronic decoding CNN. (b) The blind testing results of the system. The information hiding diffractive camera transforms various input handwritten digits into ordinary-looking "dummy" digit "8" (images shown after normalization); the electronic decoding CNN accurately reconstructs the original input handwritten digits. (c) Phase modulation patterns of the optimized diffractive layers of the information hiding camera.

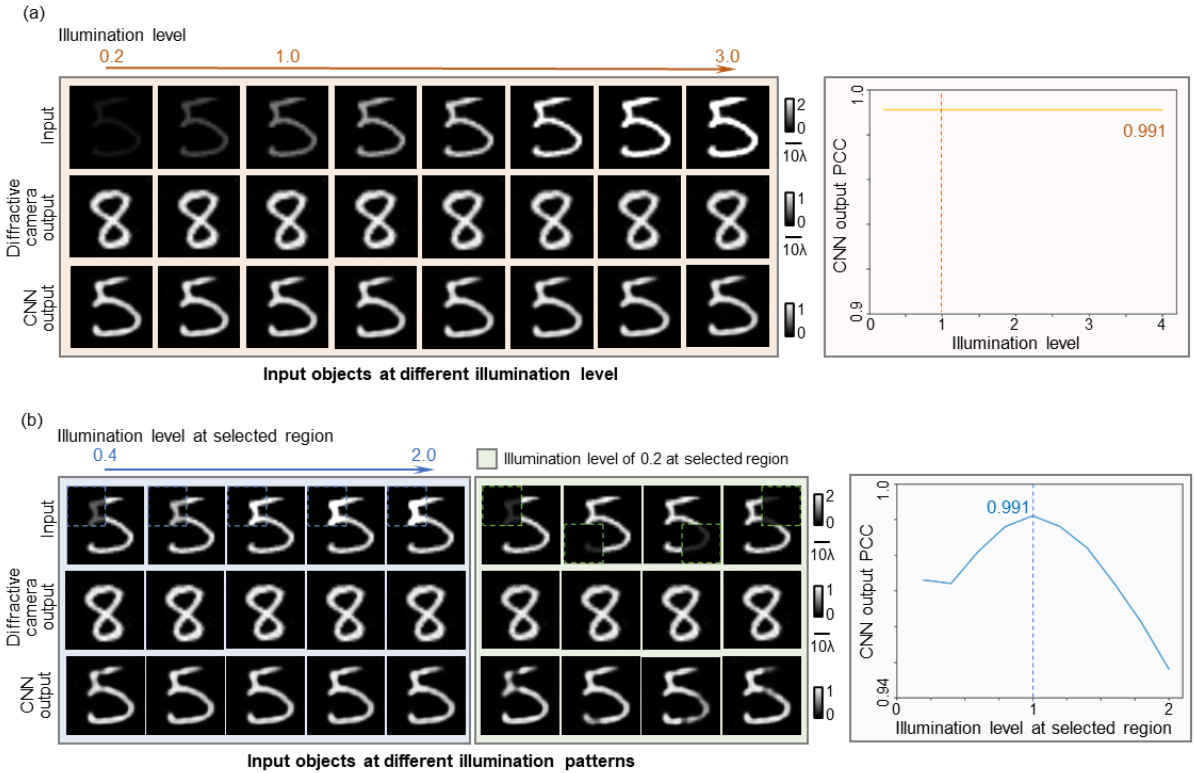

**Figure 3. Evaluation of the information hiding– decoding system trained with normalization (shown in Fig. 2) under varying illumination conditions.** (a) The blind testing examples and the quantitative PCC analysis under uniformly varying illumination intensities. (b) The blind testing examples and the quantitative PCC analysis under spatially non-uniform illumination patterns. All the images of the diffractive camera output are shown after normalization.

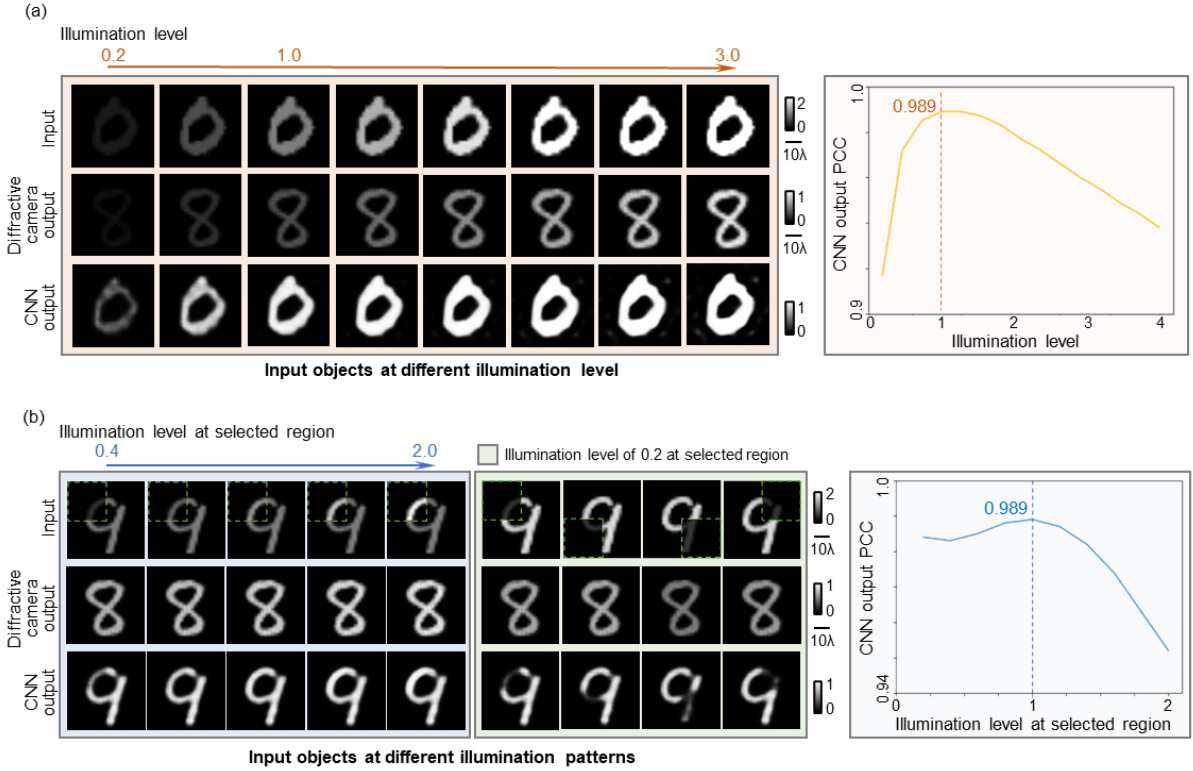

**Figure 4. Evaluation of the information hiding– decoding system trained without normalization (shown in Supplementary Fig. 1) under varying illumination conditions.** (a) The blind testing examples and the quantitative PCC analysis under uniformly varying illumination intensities. (b) The blind testing examples and the quantitative PCC analysis under spatially non-uniform illumination patterns.

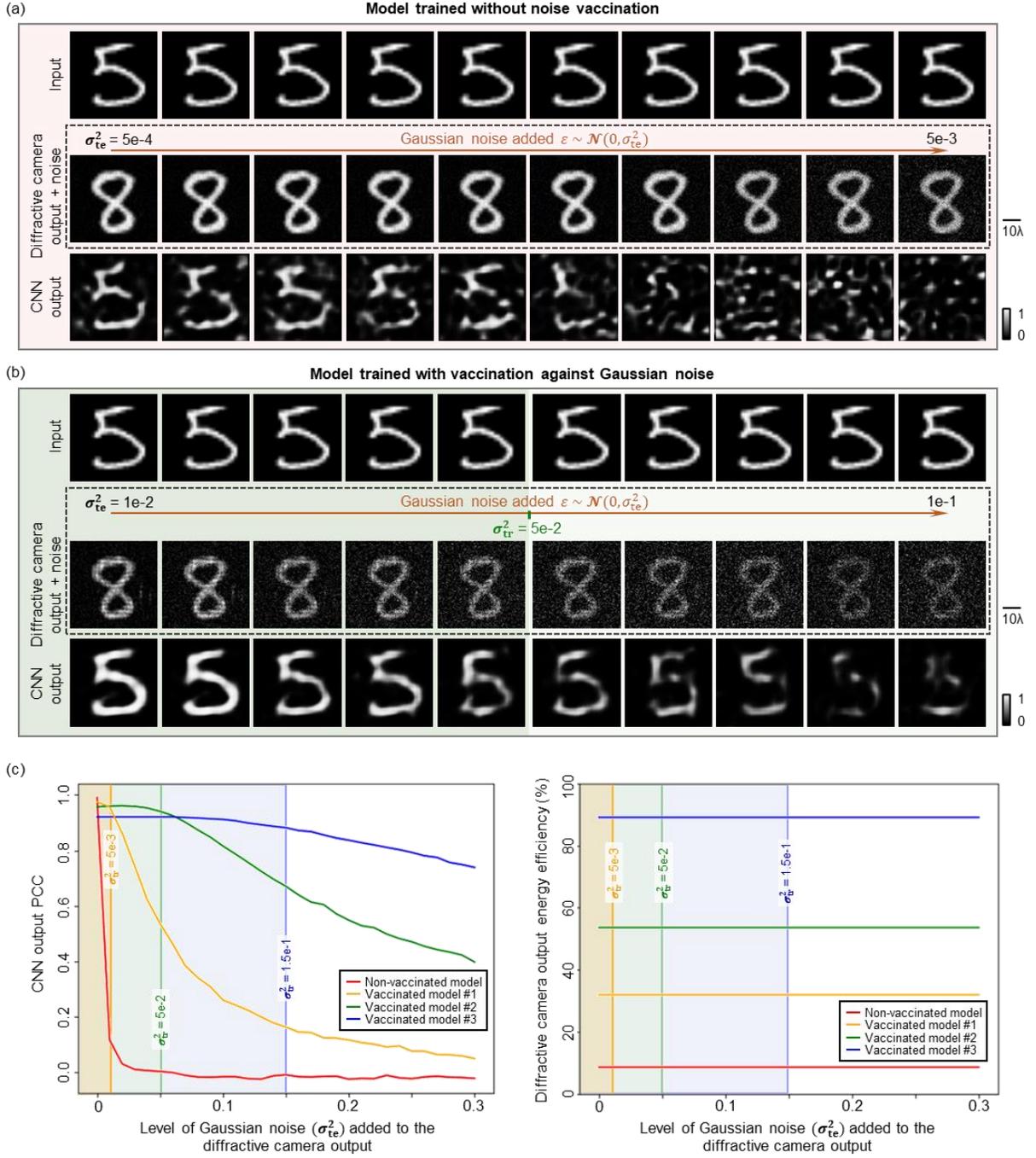

**Figure 5. Evaluation of the information hiding–decoding system performance under noisy conditions.** (a) The blind testing examples of the non-vaccinated model (shown in Fig. 2) under different noise levels. (b) The blind testing examples of the model trained with noise vaccination under different noise levels. (c) Quantitative analysis of the CNN reconstruction PCC values and the diffractive camera output energy efficiency - comparing models trained with vaccination against different noise levels. All the images of the diffractive camera output are shown after normalization.

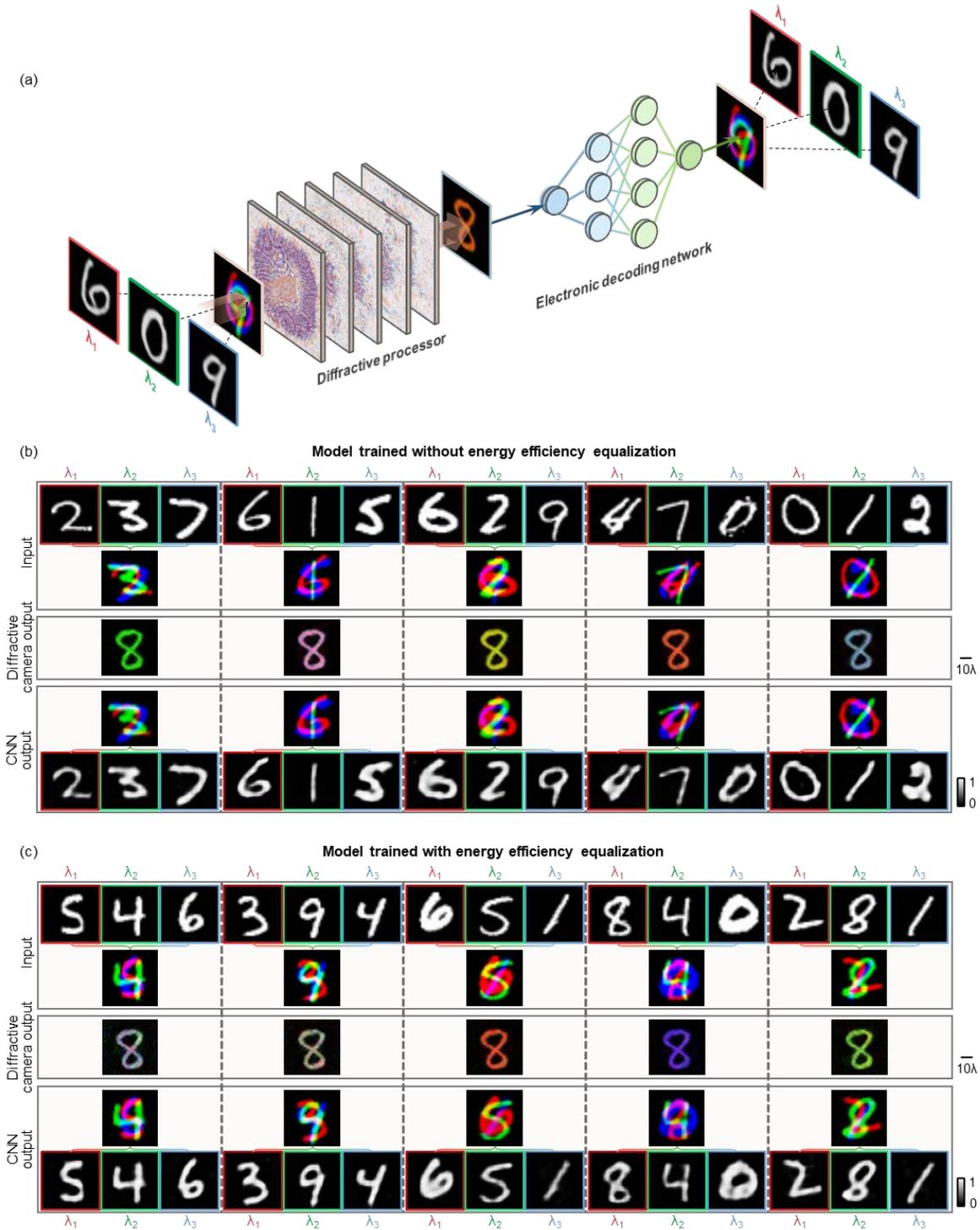

**Figure 6. Multi-spectral all-optical information hiding – decoding system**. (a) Design of the multi-spectral information hiding diffractive camera with an electronic decoding neural network. The information hiding diffractive camera conceals multiple independent images, each carried by a distinct wavelength channel, into colorful but ordinary-looking "dummy" output images, which are subsequently

decoded by a jointly trained electronic neural network. (b) The blind testing examples of the multi-spectral diffractive camera model trained *without* the energy efficiency equalization loss. (c) The blind testing examples of the multi-spectral model trained *with* the energy efficiency equalization loss. Also see Fig. 7. All the images of the diffractive camera output are shown after normalization.

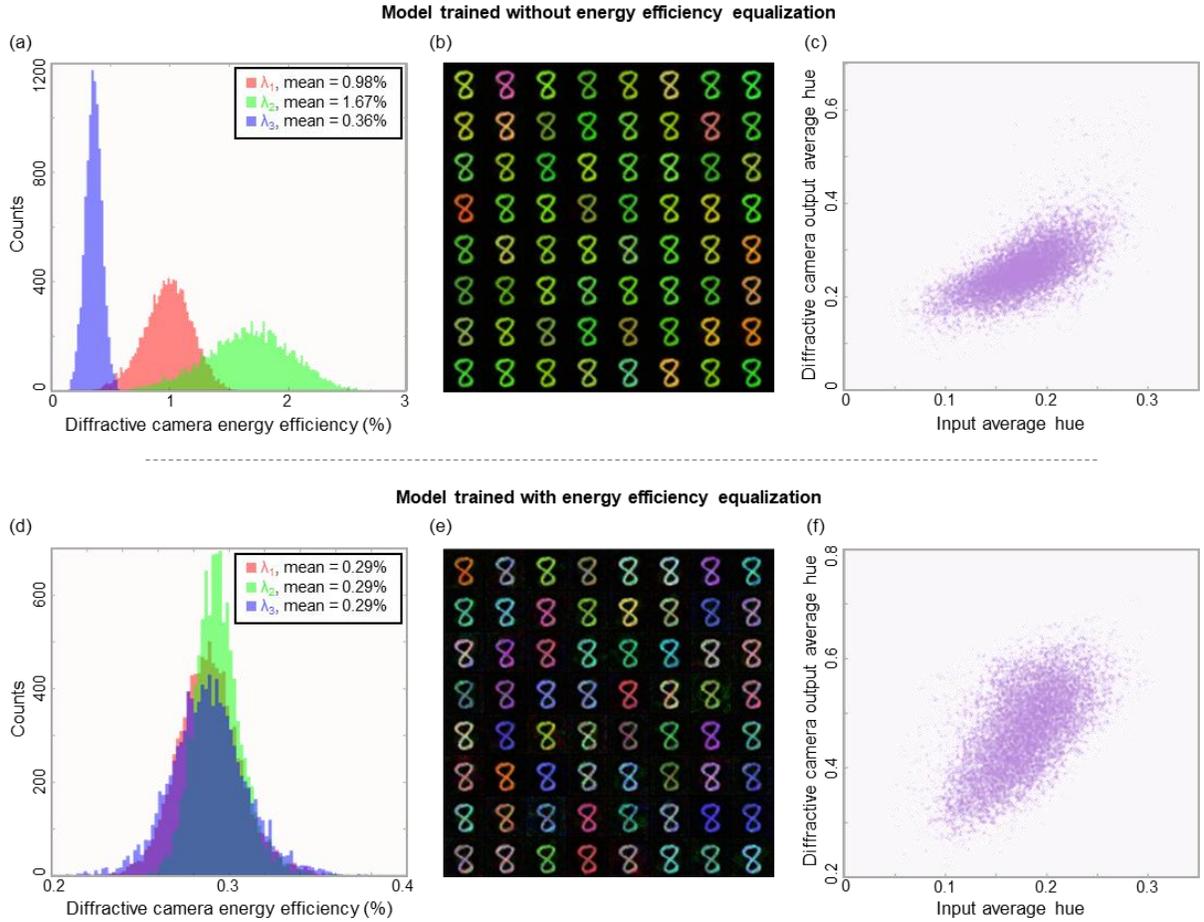

**Figure 7. Quantitative analysis of the multi-spectral all-optical information hiding camera**. (a) Histograms depicting the diffractive camera energy efficiency at each distinct wavelength, analyzed over 10,000 blind testing examples from the model trained without the energy efficiency equalization loss. (b) A set of randomly selected examples of the information hiding diffractive camera output. (c) A scatter plot comparing the average hue of the input and output images of the diffractive camera model across 10,000 test images. (d, e, f), The same set of analyses performed on the model trained with the energy efficiency equalization loss. All the images of the diffractive camera output are shown after normalization.

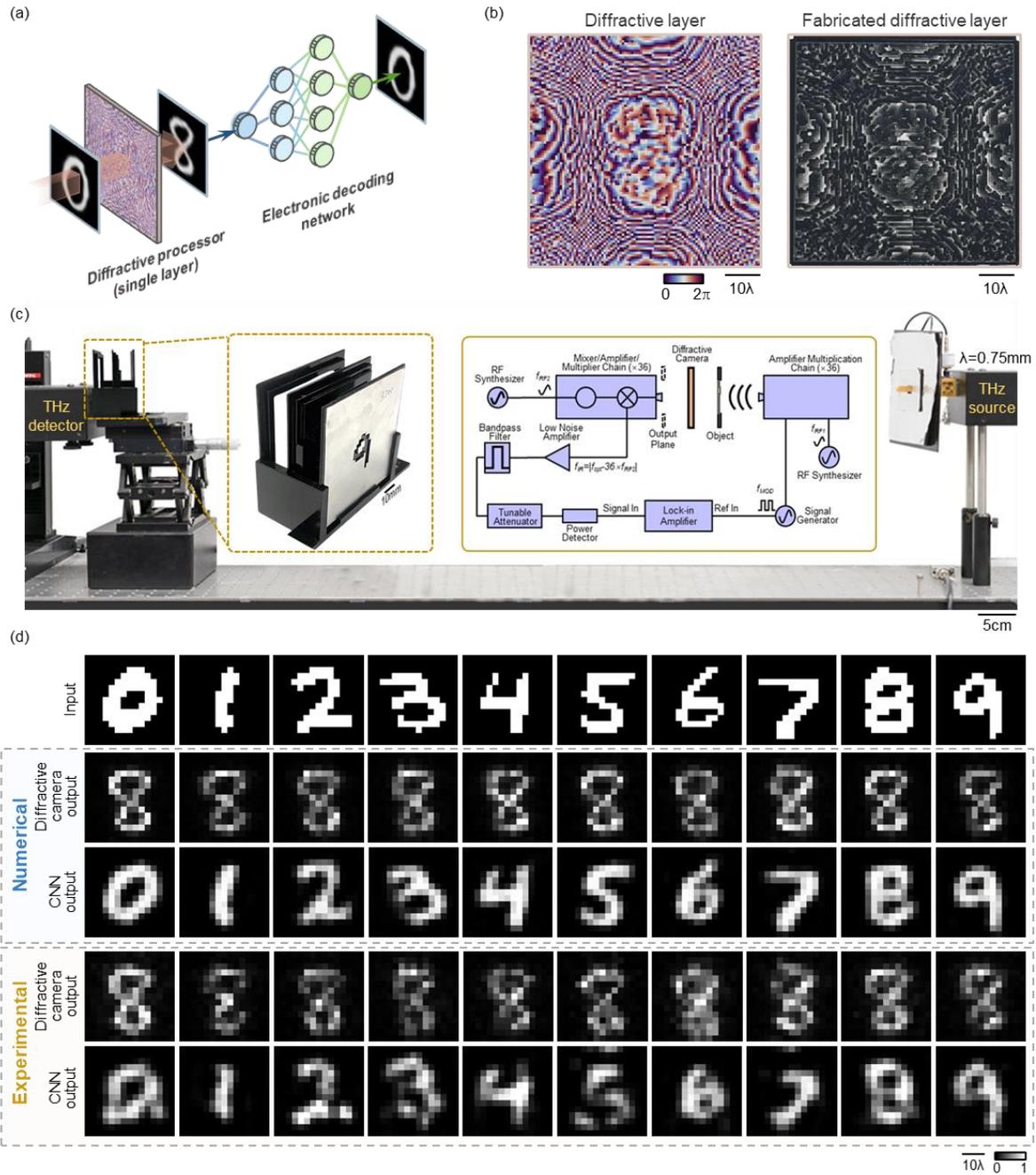

**Figure 8. Experimental validation of the all-optical information hiding diffractive camera with electronic decoding.** (a) Design schematic of a single-layer information hiding diffractive camera integrated with a jointly trained electronic decoding CNN. (b) Converged phase modulation pattern and the corresponding fabricated diffractive layer. (c) Schematic of the experimental setup under THz radiation. (d) Experimental results showing the information hiding diffractive camera measurements and the subsequent CNN reconstructions, alongside the numerical simulation results reported for comparison. All the images of the diffractive camera output are shown after normalization.